\documentclass[11pt]{article}
\usepackage{moriond,epsfig}

\def\MyPhoto{\psfig{figure=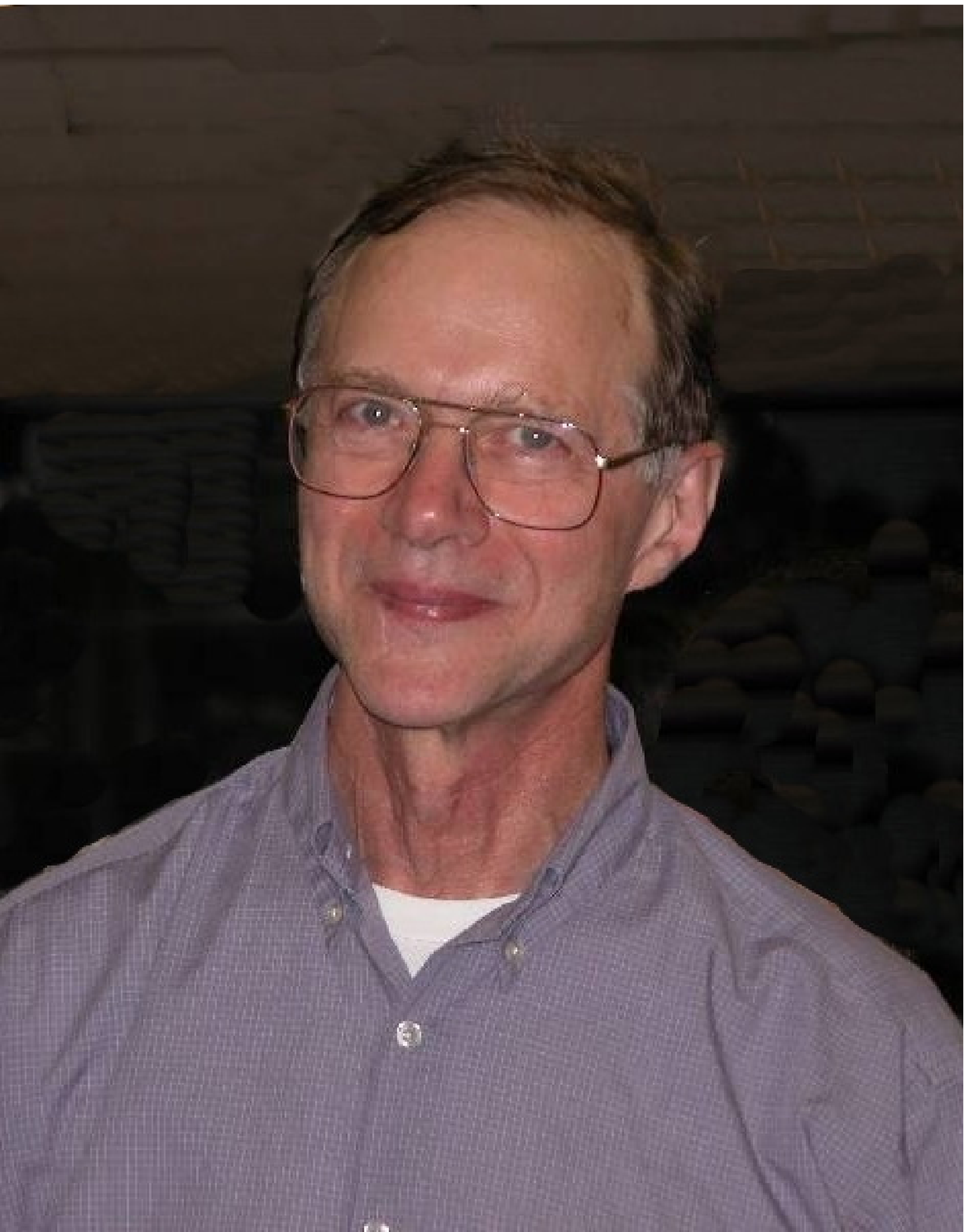,width=35mm}}

\def\Journal#1#2#3#4{{#1} {\bf #2}, #3 (#4)}

\def\PLB{{\em Phys. Lett.}  B}
\def\PRL{\em Phys. Rev. Lett.}
\def\PRD{{\em Phys. Rev.} D}

\def\ra{\rightarrow}

\def\be{\begin{equation}}
\def\ee{\end{equation}}
\def\bea{\begin{eqnarray}}
\def\eea{\end{eqnarray}}

\newcommand\Strut{\rule[-1em]{0pt}{2.6em}}  %Strut to expand table

\newcommand{\Eq}[1]{Eq.~(\ref{eq#1})}

\begin{document}
\vspace*{4cm}
\title{LEPTOGENESIS AND ITS ELECTROMAGNETIC VARIANT \footnotemark[1]}
\footnotetext[1]{FERMILAB-CONF-09-359-T. To appear in the Proceedings of the 2009 Rencontres de Moriond on Electroweak Interactions and Unified Theories.}

\author{ BORIS KAYSER}

\address{Fermilab, MS 106, P.O. Box 500, Batavia, IL 60510, USA}

\maketitle\abstracts{
We briefly explain how the present baryon-antibaryon asymmetry of the universe could have arisen through leptogenesis, and then discuss a new version of leptogenesis in which CP violation in electromagnetic decays plays the central role.}

\section{Standard Leptogenesis}
%\subsection{Producing the Hard Copy}\label{subsec:prod}

The universe presently contains $6 \times 10^{-10}$ baryons for every photon, but essentially no antibaryons. Yet, from cosmology and particle physics we believe that any initial asymmetry between the number of baryons and the number of antibaryons would have been erased shortly after the big bang. We must then understand how a universe with equal numbers of baryons and antibaryons evolved into one with many more baryons than antibaryons. Sakharov pointed out long ago that such a change from baryon-antibaryon symmetry to baryon-antibaryon asymmetry could not have occurred without a violation of CP invariance. The Standard-Model CP violation in the quark mixing matrix, observed in K and B decays, can lead only to a baryon-antibaryon asymmetry very much smaller than the one observed. However, CP violation in leptogenesis, a scenario that involves the leptons, can produce an asymmetry of the observed magnitude. 

Leptogenesis~\cite{r1}$^,$ \cite{r2} is an outgrowth of the see-saw model,~\cite{r3} the most extensively studied theory of why the neutrinos are so light. In its straightforward form, the see-saw model adds to the Standard Model (SM) only several (three, say) weak-isospin singlet, right-handed neutrinos $N_{kR}$. These are given very large Majorana masses $M_k$, and Yukawa couplings to the SM light lepton doublets and the SM Higgs doublet. Thus, in the see-saw picture, the Lagrangian is that of the SM plus 
\be
{\cal L}_{\mathrm{new}} = -\sum_{k=1}^3 \frac{M_k}{2} \overline{N_{kR}^c} N_{kR} \,- \,\sum_{j,k=1}^3 y_{jk}[\overline{\nu_{jL}}\overline{\varphi^0} - \overline{\ell_{jL}}\varphi^-] N_{kR} + h.c.~~.
\label{eq1}
\ee
Here, $\nu_{jL}$  and $\ell_{jL}$  are the members of the SM light-lepton doublet of the jÕth generation,  $\varphi^+$ and $\varphi^0$  form the SM Higgs doublet,  $y_{jk}$ is a Yukawa coupling constant, and $c$ stands for charge conjugation. 

\begin{figure}[htb]   \begin{centering}
\includegraphics[scale=1.0]{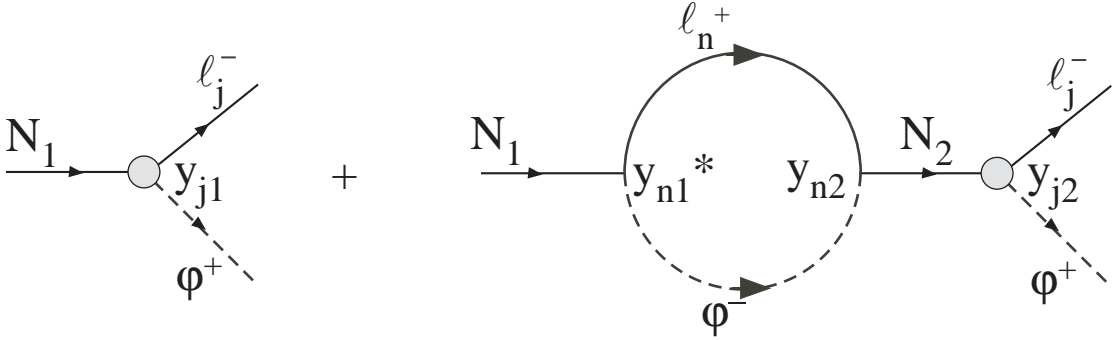}
\end{centering}
\caption{A tree diagram and a loop diagram contributing to the decay of the heavy neutrino $N_1$.}
\label{fig1}
\end{figure}

The Yukawa interaction in \Eq{1} gives rise to the decays $N_k \rightarrow \nu_j + \varphi^0$  and their CP-mirror images,  $N_k \rightarrow \overline{\nu_j} + \overline{\varphi^0}$, and to the decays $N_k \rightarrow \ell^-_j + \varphi^+$  and their CP-mirror images,  $N_k \rightarrow \ell^+_j + \varphi^-$. If there are CP-violating phases in the Yukawa coupling matrix $y$, then interferences between tree and loop diagrams will lead to CP-violating differences between the rates for CP-mirror-image N decays. For example, interference between the diagrams shown in Fig.~\ref{fig1} leads to the CP-violating difference
\be
\Gamma(N_1 \rightarrow \ell^-_j + \varphi^+) - \Gamma(N_1 \rightarrow \ell^+_j + \varphi^-) \propto \Im(y^*_{j1}y^*_{n1}y_{j2}y_{n2})~~.
\label{eq2}
\ee

Even if the heavy neutrinos $N$ are too massive to be produced and studied at present-day accelerators, they would have been present just after the hot big bang. If their early-universe decays involved CP asymmetries such as the one in \Eq{2}, and occurred out of
equilibrium, then these decays would have left the universe with unequal numbers of leptons and antileptons. This would have been Step One of leptogenesis---a two-step scenario. In Step Two, the nonperturbative SM ``sphaleron'' process would have acted. The sphaleron process does not conserve baryon number $B$, defined as the total number of baryons minus the total number of antibaryons, or lepton number $L$, defined similarly. However, this process does conserve $B-L$ . Starting from the initial state produced by Step One, with initial values $B_i = 0$  but $L_i \neq 0$, the sphaleron process would have yielded a final state with final values $B_f \simeq -\frac{1}{3}L_i$ and $L_f \simeq \frac{2}{3} L_i \simeq -2B_f$. Perhaps this two-step scenario is how the universe came to have a non-vanishing baryon number. 

In the simplest picture, the leptonic asymmetry produced by Step One comes from decay of the lightest $N$, which we call $N_1$, and the final lepton flavors, $e,\; \mu$, and $\tau$, may be treated identically. Then, summing over the final lepton flavors, the CP-violating asymmetry is
\bea
\epsilon &\equiv& \frac{\Gamma(N_1\ra L\phi) - \Gamma(N_1\ra \bar{L}\bar{\phi})}{\Gamma(N_1\ra L\phi) + \Gamma(N_1\ra \bar{L}\bar{\phi})}  \\
&=& \frac{1}{8\pi} \frac{1}{(y^\dagger y)_{11}} \sum_m \Im[\{( y^\dagger y)_{1m}\}^2] K \left( \frac{M^2_m}{M^2_1} \right)~~.  \nonumber
\label{eq3}
\eea
Here, $L$ stands for all the light lepton doublets, $\phi$ is the Higgs doublet, and $K$ is a kinematical function which is of order unity so long as  $M_m / M_1$ is not near unity or extremely large. Disregarding the matrix structure of $y$, we see that
\be
\epsilon \sim y^2 / 10 ~~ .
\label{eq4}
\ee
To explain the present ratio of baryons to photons, $\sim 10^{-9}$, we require that 
$\epsilon \sim  10^{-6}$. Thus, we must have $y^2 \sim 10^{-5}$.

In addition to giving rise to the early-universe decays of the heavy neutrinos, the Yukawa interaction in the see-saw Lagrangian of \Eq{1} plays another role. As the universe cools through the electroweak phase transition (which occurs after the heavy neutrinos have decayed), the neutral Higgs field develops its present vacuum expectation value, $\langle\varphi^0 \rangle _0 \equiv \nu \simeq 175\,$GeV. Correspondingly, the Yukawa term $\overline{\nu_L}\, y \, \overline{\varphi^0}\, N_R$ (in matrix notation) in \Eq{1} develops a piece $\overline{\nu_L}\, (y\nu)  N_R = \overline{\nu_L} M^\dagger_D N_R$, where  $M_D \equiv (y\nu)^\dagger$ is a 3 $\times$ 3 Dirac mass matrix for the neutrinos. In the see-saw model, it is assumed that, since no symmetry prevents the Majorana masses $M_k$ from being very large, they {\em are} very large. That is, $M_N \gg M_D$, where $M_N$ is the diagonal matrix whose diagonal elements are the masses $M_k$. As is well known, the mass matrix for the familiar light neutrinos, $M_\nu$, is then given by the Òsee-saw relationÓ,
\be
M_\nu \simeq -M^T_D  \frac{1}{M_N} M_D ~~ .
\label{eq5}
\ee

Disregarding the details of matrix structure, we see from this relation and  $M_D \equiv (y\nu)^\dagger$ that the light neutrino masses, empirically of order 0.1 eV, must be of order $(y\nu)^2 / M_N$. Now, we saw previously that successful leptogenesis requires that $y^2 \sim 10^{-5}$. From this requirement and the demand that $(y\nu)^2 / M_N\sim 0.1\,$eV, it follows that $M_N$ must be of order $10^9$ GeV. That is, the requirement that the Yukawa interaction in the see-saw picture give rise both to successful leptogenesis and to light neutrino masses of the observed size leads to heavy neutrinos far beyond the range of the LHC. 

In a basis where the charged lepton mass matrix is diagonal, the Yukawa coupling matrix $y$ of \Eq{1} is the only source of CP violation among the leptons. If $y$ contains CP-violating phases that drive leptogenesis, then these phases will also appear in $M_D \equiv (y\nu)^\dagger$. Thus, in general, CP-violating phases will also appear in the light neutrino mass matrix $M_\nu$, which is related to $M_D$ through the see-saw relation, \Eq{5}. Consequently, in general, CP-violating phases will also appear in the light neutrino mixing matrix $U$, which is just the matrix that diagonalizes $M_\nu$. Since CP-violating phases in $U$ lead to CP violation in light neutrino oscillation and in neutrinoless double beta decay, we expect CP violation in these phenomena if leptogenesis is indeed the explanation of the baryon-antibaryon asymmetry of the universe.

\section{Electromagnetic Leptogenesis}

Let us turn now to a new variant of leptogenesis---electromagnetic leptogenesis.~\cite{r4} Suppose that new physics at a high mass scale $\Lambda > M_N$ leads to ``electromagnetic'' $N$ decays,
\be
N \ra L + \phi + (\gamma \; \mathrm{or} \; Z \; \mathrm{or} \; W) ~~,
\label{eq6}
\ee
which yield an electroweak gauge boson in addition to the particles emitted in the $N$ decays of standard leptogenesis. Could CP violation in such electromagnetic decays produce a successful alternative to standard leptogenesis? If so, could this alternative be successful even if the $N$ masses are in the TeV range, rather than $\sim 10^9$ GeV, so that the heavy neutrinos are within range of the LHC? To explore these questions, we assume that the new physics leads to the dimension-six effective ``electromagnetic'' interaction
\be
-{\cal L}_{\mathrm{``EM''}} = \frac{1}{\Lambda^2} \sum_{j,k=1}^3 \overline{L_{jL}}\, \sigma^{\alpha \beta} \, [\lambda_{jk} B_{\alpha \beta} + \tilde{\lambda}_{jk} \vec{\tau} \cdot \vec{W}_{\alpha \beta} ] \, \bar{\phi} N_{kR} + \mathrm{h. c.} ~~.
\label{eq7}
\ee
Here, $\lambda_{jk}$ and $\tilde{\lambda}_{jk}$ are dimensionless complex coupling constants, $L_{jL}$ is the SM light lepton doublet of the $j$th generation, $\phi$ is the SM Higgs doublet, $N_k$ is one of the three isospin-singlet heavy neutrinos, $\vec{\tau}$ is the vector of Pauli matrices, and $B_{\alpha \beta}$ and $\vec{W}_{\alpha \beta}$ are the usual isosinglet and isotriplet SM field-strength tensors.

\begin{figure}[htb]        \begin{centering}
\includegraphics[scale=0.9]{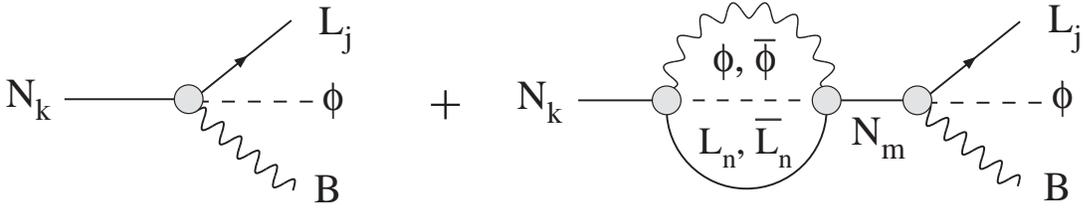}
\end{centering}
\caption{The tree diagram and a loop diagram contributing to the ``electromagnetic'' $N$ decay $N_k \ra L_j \phi B$.}
\label{fig2}
\end{figure}

If there are CP-violating phases in the coupling matrices $\lambda$ and $\tilde{\lambda}$, then, just as in standard leptogenesis, tree-loop interferences will lead to CP-violating differences between the rates for CP-mirror-image $N$ decays. For example, the interference between the tree and loop diagrams in Fig.~\ref{fig2} can lead to $\Gamma(N_k \ra L_j \phi B) \neq \Gamma(N_k \ra \bar{L_j} \bar{\phi} B)$, where $B$ is the isosinglet SM gauge boson. Explicit calculation shows that, for suitable values of the parameters, electromagnetic and standard leptogenesis can yield similar CP-violating asymmetries $\epsilon$.

\begin{figure}[htb]    \begin{centering}
\includegraphics[scale=0.8]{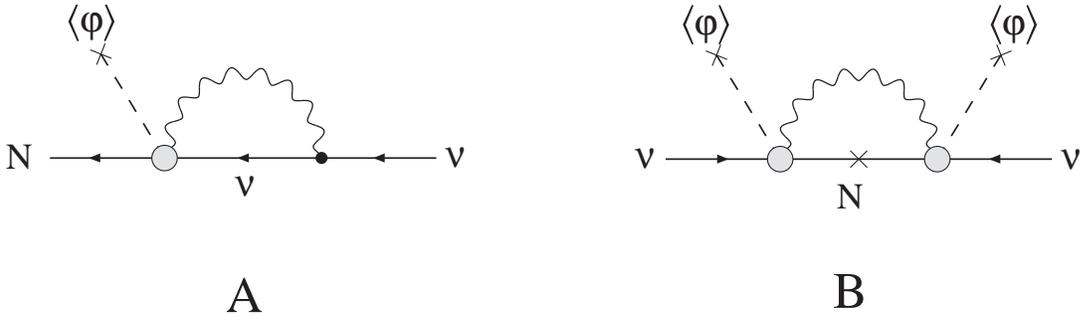}
\end{centering}
\caption{Diagrams through which the new effective EM interactions contribute to the light neutrino masses.}
\label{fig3}
\end{figure}

Before discussing this more quantitatively, we note that, once the universe cools through the electroweak phase transition and $\varphi^0$ acquires its vacuum expectation value, the new ``EM'' interactions of \Eq{7} contribute to the light neutrino masses through the diagrams in Fig.~\ref{fig3}.
In these diagrams, the blob is the new EM interaction. Diagram A, in which $g^\prime$ is the SM $U(1)$ gauge coupling constant, is a Dirac mass term that will contribute to the light neutrino mass via the see-saw mechanism. Diagram B is a Majorana mass term that will contribute to the light neutrino mass directly.

Table \ref{t1} compares the $N_1$ decay rate $\Gamma_1$, the CP asymmetry $\epsilon$ in this decay, and the light neutrino masses arising from the EM interactions of \Eq{7} to their standard-leptogenesis counterparts that arise from the Yukawa couplings of \Eq{1}. 
\begin{table}[htb]
\caption{Comparison between standard and electromagnetic leptogenesis. The mass terms $m^A_\nu$ and $m^B_\nu$ correspond, respectively, to the diagrams A and B in Fig.~\ref{fig3}.\label{t1}}
\vspace{0.4cm}
\begin{center}
\begin{tabular}{|c|c|}
\hline 
\Strut    {\bf Standard} & {\bf Electromagnetic}  \\ 
\hline 
\Strut $\Gamma_1 = \frac{1}{8\pi} (y^\dagger y)_{11} M_1$  &  $\Gamma_1 = \frac{1}{2\pi} (\lambda^\dagger \lambda)_{11} M_1 \left( \frac{M_1^2}{8\pi \Lambda^2} \right) ^2$
\\  \hline
\Strut   $\epsilon \sim \frac{1}{8\pi}  \frac{\Im(y^\dagger y)_{1m}^2}{(y^\dagger y)_{11}} \frac{ M_1}{M_m} $ &$ \epsilon \sim \frac{1}{2\pi} \frac{\Im (\lambda^\dagger \lambda)^2_{1m}}{(\lambda^\dagger\lambda)_{11}}  \frac{M_1}{M_m} \left( \frac{M_1^2}{8\pi \Lambda^2} \right) ^2$
\\   \hline 
\Strut $m_\nu \sim y^*\, M^{-1}_N y^\dagger \langle\varphi\rangle^2$  & $m^A_\nu \sim \lambda^T M_N^{-1} \lambda \langle \varphi \rangle^2 \left( \frac{g^\prime}{16\pi^2}\right)^2 $
\\ 
\Strut  & $m^B_\nu \sim \frac{\lambda^ T M_N \lambda}{\Lambda^2} \langle \varphi \rangle^2 \frac{1}{16\pi^2}$
\\   \hline
\end{tabular}
\end{center}
\end{table}
The EM column of this table includes only contributions from the couplings to the isosinglet gauge boson $B$, and among the loop diagrams only the self-energy diagram of Fig.~\ref{fig2}, but the contributions of the couplings to the isotriplet gauge boson $W$, and those involving vertex-correction loop diagrams, are expected to be of similar magnitude.

From Table \ref{t1}, we see that apart from the suppression factor $(M^2_1 / 8\pi \Lambda^2)^2$, the CP asymmetry $\epsilon$ produced by electromagnetic leptogenesis is very similar to that produced by standard (Yukawa) leptogenesis, with the coupling matrix $\lambda$ in the former playing the role of $y$ in the latter. Thus, if standard leptogenesis can successfully explain the baryon-antibaryon asymmetry of the universe, so can EM leptogenesis. However, from Table \ref{t1} we also see that the neutrino masses stemming from the EM couplings $\lambda$ are not so different from those arising from the Yukawa couplings $y$. Thus, if standard leptogenesis cannot be compatible with both the observed baryon asymmetry of the universe and the observed rough masses of the light neutrinos unless the heavy neutrinos $N$ have masses far above the range accessible to the LHC, the same is true of EM leptogenesis.

To illustrate, let us assume that the Yukawa couplings $y$ are negligible, so that $N$ decays and the light neutrino masses are both dominated by the EM couplings $\lambda$. Disregarding the matrix structure of $\lambda$, taking $\lambda \sim 35$ and $\Lambda \sim 10\,M_2 \sim 20\, M_1$, we find from Table \ref{t1} with $m=2$ that $\epsilon \sim 10^{-6}$. This value of the CP asymmetry successfully accounts for the present ratio of baryons to photons. If we now take, in particular, $M_1 \sim 5 \times 10^{12}\,$GeV, and disregard the matrix structure of $M_N$, we find that $m^A_\nu \sim 0.03\,$eV and $m^B_\nu \sim 0.1\,$eV. Thus, the EM interaction of \Eq{7} can indeed account for both the cosmic baryon asymmetry and the light neutrino masses, but only if the $N$ masses are very large.

As noted earlier, if standard leptogenesis occurred in the early universe, then one expects CP violation in light neutrino oscillation today. We now see that if electromagnetic, rather than standard, leptogenesis occurred, then we still expect CP violation in light neutrino oscillation. As we have observed, the new EM couplings that would drive EM leptogenesis also lead to neutrino masses. If leptogenesis is dominated by these couplings, then the neutrino masses probably are as well. CP-violating phases in the couplings will lead to CP-violating phases in the light neutrino mass matrices (see Table \ref{t1}). In turn, the latter phases will lead to CP-violating phases in the light neutrino mixing matrix, and consequently to CP violation in light neutrino oscillation. Thus, whether the baryon asymmetry of the universe is due to standard leptogenesis or to its electromagnetic variant, we expect to see CP violation in light neutrino oscillation, and the search for this CP violation as a test of the general hypothesis of leptogenesis is very strongly motivated.

%%%%%%%%%%%%%%%%%%%%%%%%%%%%%%%%%%%%%%%%

\section*{Acknowledgments}
The author would like to thank Nicole Bell and Sandy Law for a fruitful and enjoyable collaboration. He would also like to thank the Aspen Center for Physics for its hospitality during the time that this collaboration was started, and during the time when this written version of a talk at the LXIVth Rencontres de Moriond was begun.


\section*{References}
\begin{thebibliography}{99}

\bibitem{r1} M. Fukugita and T.Yanagida, \Journal{\PLB}{174}{45}{1986}.

\bibitem{r2} For reviews of leptogenesis, see S. Davidson, E. Nardi, and Y. Nir, \Journal{\em Phys. Rept.}{466}{105}{2008}, and W. Buchm\"{u}ller, R. Peccei, and T. Yanagida, \Journal{\em Ann. Rev. Nucl. Part. Sci.}{55}{311}{2005}.

\bibitem{r3}  M. Gell-Mann, P. Ramond, and R. Slansky, in: {\it Supergravity}, eds. D. Freedman and P. van Nieuwenhuizen (North Holland, Amsterdam, 1979) p. 315;
\\
T. Yanagida, in: {\it Proceedings of the Workshop on Unified Theory and Baryon Number in the Universe,} eds. O. Sawada and A. Sugamoto (KEK, Tsukuba, Japan, 1979);
\\
R. Mohapatra and G. Senjanovic, \Journal{\PRL}{44}{912}{1980} and \Journal{\PRD}{23}{165}{1981};
\\
P. Minkowski, \Journal{\PLB}{67}{421}{1977}.

\bibitem{r4} N. Bell, B. Kayser, and S. Law, \Journal{\PRD}{78}{085024}{2008}. For a detailed description of this scenario, please see this paper.

\end{thebibliography}
\end{document}